# Smartphone app with usage of AR technologies – SolAR System

*G.O.Shchur, N.B.Shakhovska*

*Lviv Polytechnic National University, Lviv, Ukraine; e-mail: shchurglib@gmail.com*

*Abstract*. The article describes the AR mobile system for Sun system simulation. The main characteristics of AR systems architecture are given. The differences between tracking and without tracking technics are underlined. The architecture of the system of use of complemented reality for the study of astronomy is described. The features of the system and the principles of its work are determined.

## INTRODUCTION

The current level of development of computer technology makes it possible to significantly improve the performance of routine, monotonous works. Approaches to education and self-education have also significantly changed. One of the methods of automation and alternative method of information is the application of computer vision technologies in various fields. Recently, it is a very popular trend, which is constantly evolving and being implemented in different spheres of life.

Computer vision is the theory and technology of creating machines that can see. As a scientific discipline, computer vision refers to artificial intelligence systems that receive information from video data.

Added reality is a mixed reality that is created with the help of additional graphic elements displayed on the screen of the device. Man receives the bulk of information about the outside world through the visual channel and effectively processes, analyzes, and interprets the information received from the outside. Therefore, the question arose as soon as possible to implement a similar video processing system for computing. Technologies of computer vision and added reality are used in quite popular areas of science and technology, such as production equipment control, mobile control systems, biomedical research, product quality improvement, etc.

## THE LITERATURE OVERVIEW

### The concept of Augmented Reality

Virtual reality (VR, virtual reality) is an illusion of reality created using computer systems that provide visual, sound and other sensations. Unlike virtual reality, which requires full immersion in the virtual environment, the augmented reality uses the environment around us and simply imposes on it a certain particle of virtual information, such as graphics, sounds, and responses to touches. Because virtual and real worlds coexist harmoniously, users with the added experience of reality have the opportunity to try a whole new, improved world where virtual information is used as an additional useful tool that provides assistance in everyday activities [1-3].

Added reality (in English translation augmented reality, AR) is an addition to the physical world with the help of digital data in real time. It is provided by computer devices [4-7].

It is important to understand the differences between the complemented reality and the mixed reality. In a broad sense, complemented reality is the process of viewing the real world and virtual objects at the same time, where virtual information is superimposed, aligned and integrated into the physical world. In the literature on human-machine interaction, the added reality is in a continuous range of interfaces from "reality" to the virtual reality of "full immersion" (Fig.1)

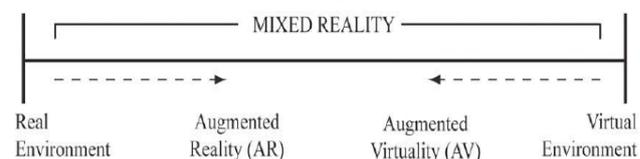

**Fig. 1.** Varieties of mixed reality

### Examples of the usage of Augmented Reality

One of the best examples of complemented reality is the Pokemon Go game. According to the rules of the game, the player must move around the real world, while his virtual "hero" travels through a virtual map through the use of GPS. If a Pokemon is next to the player, the game announces it and suggests to find it using a smartphone or tablet.

The image of the Pokemon is superimposed on the image of the surrounding world, obtained from the camera. In this case, the type of creature is calculated, depending on the information on the terrain. For example, water Pokemon will meet near rivers and fountains [8].

Another example of using add-on reality is the Stories feature in Instagram or Snapchat applications to publish their photos and videos with masks. Special algorithms of the application recognize the contours of a person's face when the camera is turned on, and impose special filter masks on it in real-time: a brush ear, glasses, colored stars or hearts.

AR uses the environment around us and simply overlays a certain piece of virtual information, such as graphics, sounds, and responses to touch.

**Types of Augmented Reality technologies**

There are several different technologies used to run AR [9-12].

The first type is marker-based. Sometimes it's also called as image recognition. This type of technology uses a camera and a special passive visual marker, such as a QR code that shows the programmed result only when the camera sensor reads it. This way it is possible to distinguish virtual objects from the real world.

The second type is a non-marker supplemented reality. Sometimes it is also called coordinate, or GPS-oriented. To provide data about your location, it can use a global positioning system, a digital compass, a speed sensor, or an accelerometer that is equipped with your device.

The most common cases of using a non-marker supplemented reality are the designation of directions, the search for the right places, such as a cafe or a museum, or in app-targeted locations.

The third type is an added reality based on projection. It works by designing light forms on physical surfaces. Special applications help to interact with a person and a projection, identifying the moments of the human touch to the projected light.

The fourth type is the addition of the VIO (Visual Inertial Odometry) based reality, a technology that helps track position and navigate through sensors and cameras, making it possible to create a precise 3D model of space around the device, update it in real time, define its position, transmit this data to all applications and apply additional layers over it.

The features of this technology are really unique: you can measure distances, insert various objects into the interior and interact with them. VIO promises to be the most promising technology in the field of complemented reality, and now it is used by such well-known companies as Google and Apple.

Odometer is a way of estimating movement using data from motion sensors.

**Areas for use of AR**

The possibilities of using AR-technologies are virtually limitless, augmented reality can be applied in almost all aspects of our lives. It will change the way we communicate, consume information and do business. This determines the need for competence, not only AR technology, but also all others [13-15].

*Education.* Augmented reality can make learning more interactive through special tours in which young researchers can view various objects while the teacher talks about them.

*Medicine.* Vipaar company has offered a solution that combines telemedicine and AR. With the help of Google Glass and the Vipaar add-ons, surgeons can assist their colleagues at a distance by designing their hands on the surgeon's glasses that are undergoing surgery.

*Aviation.* AR has long been used by military pilots. Special displays and helmets display information on fighter systems and help guide the target. AR begins to implement into the civil aviation. For example, the company Aero Glass has developed special-purpose glasses of added reality that help the pilot navigate the space, follow the route and receive additional information during the flight.

*Marketing.* Through AR technologies, brands can create more creative online advertising companies, thereby attracting additional attention to their products.

*Tourism.* It is inherent in the new generations to know the world through direct interaction. For example, the Catalan National Museum of Art began to actively use AR to better navigate intricate corridors and interactively get acquainted with exhibits.

*Design.* It was much easier to build an apartment, because instead of figuring out if a particular piece of furniture would fit you, you can virtually put it into your interior using the ARD-application Furniture Dropping.

*Shopping.* While you are in a supermarket, AR-based applications will help you navigate through a large number of rows and find the best way to get the right product. In addition, there will be an opportunity to get additional information on discounts and profitable offers.

*Games.* Perhaps there is no one who would not be in a way that would be confronted with the insanity of Pokemon Go. Someone himself was running around the city looking for Pokemon.

## FORMAL STATEMENT OF THE PROBLEM OF CONSTRUCTING AN OBJECT IN THE AUGMENTED REALITY

**Mobile add-on systems**

Mobile add-on systems include mobile apps for phones. Mobile AR means the use of various mobile interfaces for user interaction with virtual, data that complements the real world. The use of mobile phones for complemented reality has both advantages and disadvantages. Most mobile devices are now equipped with cameras, which makes the mobile phone one of the most convenient platforms for implementation of the systems of augmented reality. In addition, most cellular phones have additional built-in sensors such as accelerometers, magnetometers and GPS receivers that can improve the performance of the AR program. But, despite the rapid progress in the development of mobile phones, their computing power for complex applications is still rather small [16-20].

As a result, in many applications, client-server architecture is used when data is transferred to a remote computer that generates the computation and sends the result back to the mobile device. But with this approach, there may be a problem with limited bandwidth, and this can be critical for complex AR systems. Nevertheless, given the rapid development of mobile technology, this problem can be solved soon, which means that it will soon be possible to create applications that process data for AR locally in real time.

A successful mobile AR system, as an application, is a system that allows the user to focus on the very functionality of the system, implements the interaction of the device in a natural and socially acceptable manner, and provides the user with additional useful information. This points to the need to develop in lightweight, portable mobile devices with sufficient power for complex computing and high performance sensors for reliable tracking and recognition.

**Tracking technologies for mobile systems**

It is well known that for high-quality AR systems, in order to provide a realistic result, one needs to accurately trace the real environment for the further integration of virtual objects into it. The most common type of surveillance system for mobile systems is tracking by combining data from multiple sensors. Street systems mainly use GPS or inertial tracking methods using accelerometers, gyroscopes, compasses and other sensors, along with computer vision techniques. The GPS system provides ease of tracking, despite its low accuracy. For a more accurate assessment of the user position and its orientation, GPS is used in conjunction with different inertial sensors [21 – 23]

Thus, the user's points of interest are narrowed down, and this allows you to simplify visual tracking. In the premises, GPS has poor performance, and therefore can not be used, so only visual and inertial methods are used. The combination of these methods has its own characteristics: visual tracking achieves the best results at low frequency, and inertial sensors work better at a high frequency of motion. During slow motion, they do not produce good results due to noise and shift drift. The complementary nature of these systems leads to their common use in most hybrid systems. [14]

Some systems rely only on computer vision, but most of them are designed for work in rooms where the environment is easily controlled. When it comes to visual tracking on the street, there are external factors that greatly complicate the task. One of the most "advanced" mobile systems is Google Goggles; this system can: recognize objects of a simple form, such as bar codes or books; determine the location and direction of travel, thanks to GPS and accelerometer, which help the system determine the direction of view to narrow the point of interest.

**Simulation of Virtual Objects**

An analytical review of methods and tools for the development of complemented reality was conducted. To implement the technology of complemented reality, two major software components are required: tracking and visualization. At present, researchers and specialists have developed a large theoretical and algorithmic basis for their implementation, both in the form of separate individual components, as well as in the form of integrated programs and sets of development tools. Some components include libraries and frameworks of computer vision, engines of three-dimensional graphics, other solutions.

Tracking is a complex process associated with tracking the observer's position with respect to the surrounding environment. Optical tracking based on markers was selected for use in the study, as the most functional of the variants used for mass implementation. With this kind of tracking, the analysis of frames of the video stream coming from the camera is carried out for the presence of a special image - a marker. When a marker is successfully recognized, the matrix of transformations is calculated, which allows you to determine the position of the camera and then correctly integrate the virtual object in the real environment.

As a traction engine, the Qualcomm Vuforia development kit was used. This tool has been specifically designed for mobile devices. It is based on the algorithms of one of the leading research organizations in the area of complemented reality of the "Christian Doppler Laboratory on Added Reality".

To implement part of the technology of complemented reality associated with three-dimensional visualization, many existing solutions can be applied. During the search and analysis, the Unity development environment was selected. It has a graphical engine optimized for work on mobile devices.

**Preparation and storage of models**

The implementation of techniques to ensure realism caused the need to develop a certain process of preliminary preparation of models (Fig. 2), as well as a special structure of storage of objects (Fig. 3).

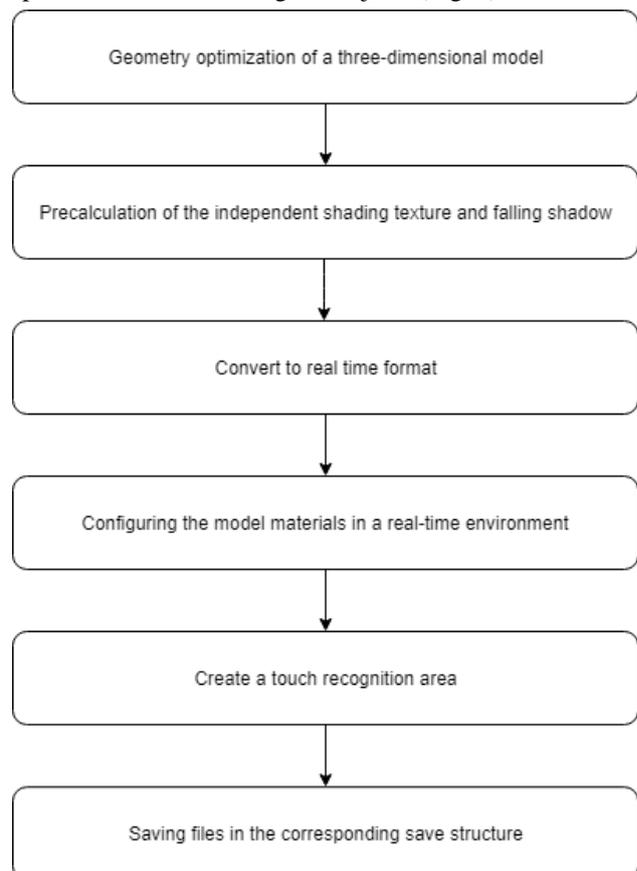

**Fig. 2.** Methodology for preparing a three-dimensional model. Analysis of methods working with control points

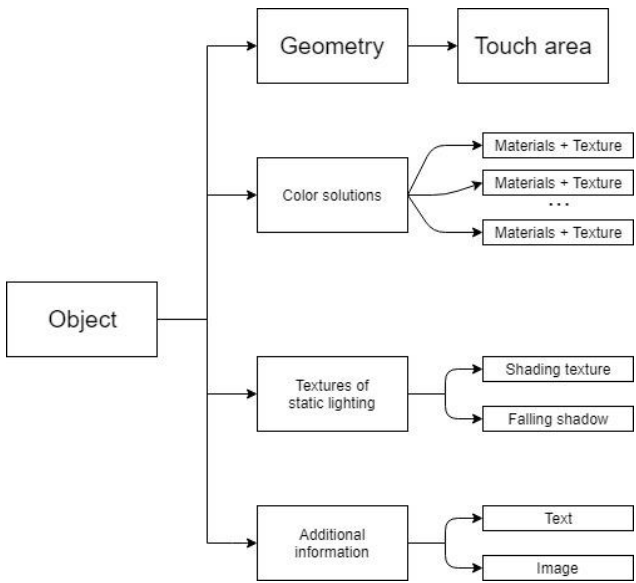

**Fig.3.** Structure of object storage

**Mechanism of manipulation of three-dimensional objects of complemented reality using signing methods of input**

In order to improve user interaction, a mechanism for manipulating three-dimensional objects using the use of sign-writing methods was developed.

As noted in studies, the task of moving objects can often be simplified to two-dimensional. And thus, in this case, the movement is carried out within the plane, along the axes X and Z, and the rotation - around its axis Y. In this case, the coordinate Y remains unchanged, and the coordinates X and Z change according to the expression:

$$\begin{bmatrix} Z_1 \\ X_1 \end{bmatrix} = \begin{bmatrix} \cos\varphi & -\sin\varphi \\ \sin\varphi & \cos\varphi \end{bmatrix} \cdot \begin{bmatrix} Z_0 \\ X_0 \end{bmatrix} + \begin{bmatrix} r_3 \\ r_1 \end{bmatrix}$$

where φ is the angle of the object around the Y axis.

Thus, we obtain a direct problem of kinematics with known solutions.

Most mobile touch screen screens are equipped with multi-touch recognition function at the same time (so-called multi-touch function). Thus it is possible to assign one gesture (simple touch and move one finger) to move the object (broadcast coordinate), and the other (touching simultaneously with two fingers) - for rotation (Fig. 4).

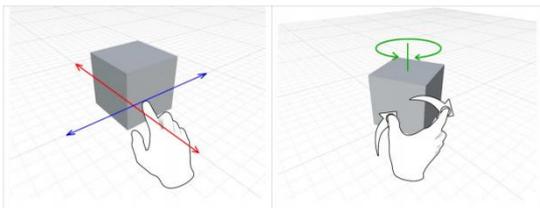

**Fig. 4.** Using the gesture "touch and move with one finger" to move the object and the gesture "touching two fingers simultaneously" - for rotation

An algorithm that handles input from the touch screen was developed (Fig. 5). When a touch screen is detected, a check is made, an incident occurred in the area of the virtual object. For this purpose a method of passage of rays is applied.

The virtual ray passes from the touch point by the user of the touch screen located on the screen plane found with the Unity virtual camera. In the case of intersection of the beam of the object, returns a positive result. If there are several virtual objects on the scene, it is necessary to determine with which of them the interaction is carried out.

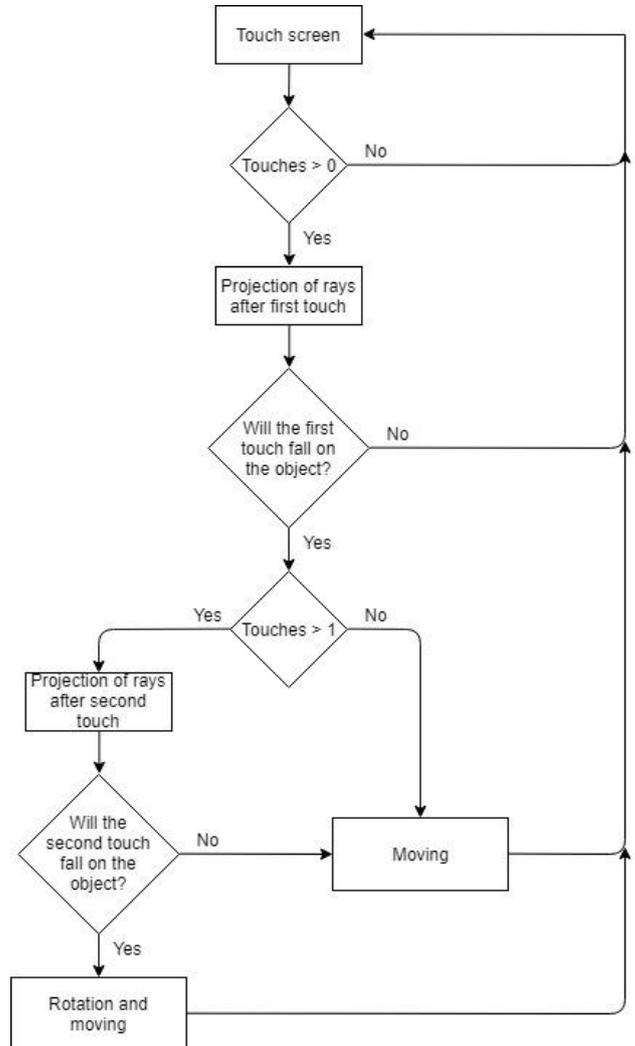

**Fig. 5.** Block diagram of object manipulation algorithm

When detecting more than one touch on one object, the rotation mode is recognized. The angle at which the object is to be rotated is determined by measuring the line of rotation through two points of contact, in comparison with the previous frame. when more than two points are detected on an object, only the first two are processed.

PRACTICAL IMPLEMENTATION AND ANALYSIS OF RESULTS

The system has two interfaces that receive input. One is the web interface for administrators or vendors to create a mall shopping mall by downloading images and matching metadata. Another Android application that customers use to send queries. To process customer requests, the procedure consists mainly of three stages, that is, the construction of the model, the coordination of functions and the generation of annotations. The first step is disabled until the last two steps are online. In the process of creating a model, administrators should send images and metadata through our Android application through the web interface. When a client picks up an image using the application and sends it to the server, spatial functions and visual functions will be extracted from the metadata and image. Upon completion of the matching, annotation functions from models and images downloaded by other clients will be transferred to the query image and displayed on the screen.

The Unity development environment is used. To design 3D objects, the MaxStAR plugin is used. The scripts used in the program are written in C #.

AForge.NET is an open-source C # framework developed for developers and researchers in the field of computer vision and artificial intelligence. The framework includes the following components:

AForge.Imaging - image processing and a set of different filters;

AForge.Vision - a set of methods and algorithms of computer vision;

AForge.Video - video stream processing;

AForge.Neuro - construction and work with neural networks;

AForge.Genetic - a set of genetic algorithms;

AFroy.Robotics - a special set of methods for use in the field of robotics.

On the basis of AForge.NET, the Gratf library was developed, which is used to build an augmented reality. It is written in C # and well tolerated on different platforms.

In addition to the above, there are a number of other cross-platform computer vision libraries. Among them is ROS (Robot Operating System) - an open source library used to create robots software. VXL, Integrating Vision Toolkit, ViSP - C ++ frameworks with a set of modules for processing and analysis of images, video stream, search patterns and objects, classifiers and many more. The list of similar libraries is quite extensive, and a significant number of them are based on OpenCV.

Unity is a gaming engine that allows the development of two-dimensional and three-dimensional computer games and other applications for all existing platforms, it works on Windows, Linux, and OS X platforms.

**Architecture of the system**

The following modules are required for the operation of this system:

1) The camera

The camera component ensures that each frame view is captured and transmitted efficiently to the tracker. The developer only initializes the camera to start and stop the broadcast. The frame of the camera is automatically converted into a hardware-dependent format and sets the desired image size.

2) Image Converter.

The format converter converts frames from a format camera to a format suitable for WebGL rendering and tracking. This conversion also involves reducing the image from the camera in various resolutions available in the converted stack frame.

3) The tracker

Component tracker contains computer vision algorithms, in order to detect and track real-world objects within the camcorder. Based on a camera image, different algorithms take care of identifying new targets or markers. A tracker can download multiple sets of data at a time and activate them. In our case, a flexible system is implemented, which can work without this module.

4) Texture sketching

This visualization module creates an image stored in the object.

5) Application code

All components listed above must be initialized in the code and three conditions must be fulfilled. For each processed frame, the object is updated and the method of playback is called. Required:

- Define objects for newly discovered targets, markers, or updated states of these elements;

- Update the logic of the program with new inbound data;

- Draw a figure with a texture imprinted on it.

6) Marker base

Create a marker object that specifies markers that will recognize the application.

**The program results**

The screenshots of the augmented system are shown.

As can be seen from Fig. 6 - 8, the system operates in a mode where there is no need to use a tracker, so a virtual object can be designed for different images.

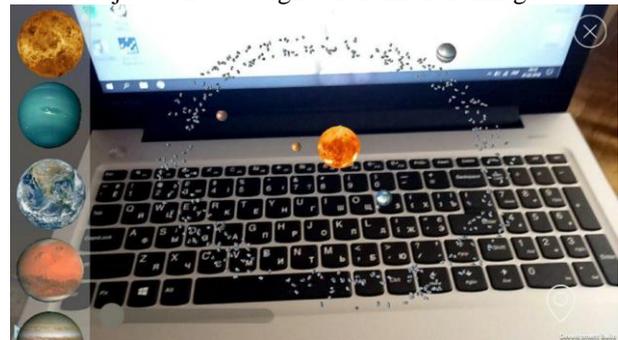

**Fig. 6**

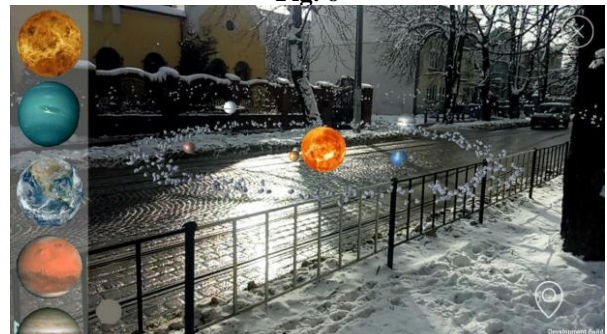

**Fig. 7**

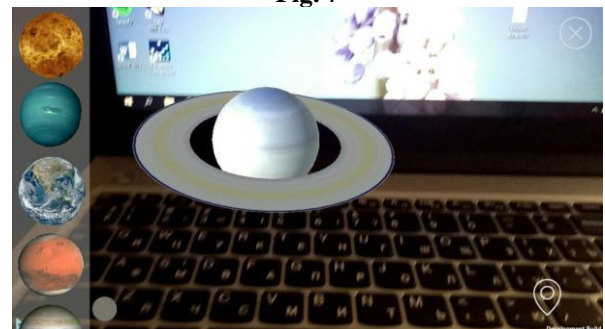

**Fig.8**

In Fig. 9 the main application window is shown.

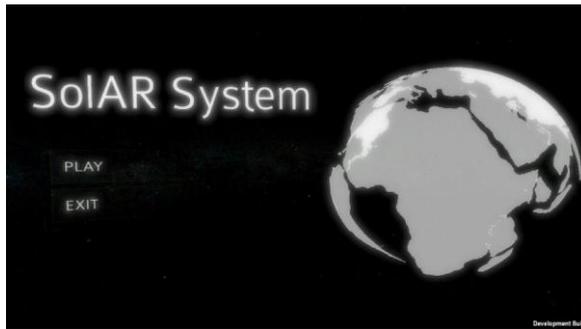

**Fig. 9.** The main window

## CONCLUSIONS

The advantages of the developed system are:
1. **The program does not need any sheets of paper with markers on it.** Our program uses the most up-to-date version of computer vision, which does not require any piece of paper. The smartphone is guided in space and understands how to properly design objects.
2. **Optimization.** At the expense of optimization it is achieved that the program does not need a large amount of resources of the smartphone, so even relatively weak smartphones will be able to run the program without problems, while not discharging the battery in a matter of minutes.
3. **Starts on smartphones with Android version starting from 4.1**. This enables a much larger number of users to use our program.